\documentstyle[aas2pp4]{article}
\newcommand{\rs}{R_*}
\newcommand{\mdot}{\dot M}
\newcommand{\MS}{M_{\odot}}
\newcommand{\Sh}{Shvartsman }
\newcommand{\Sha}{Shvartsman}
\begin{document}

\title{On Electrostatic Positron Acceleration In The \\
Accretion Flow Onto Neutron Stars}
\author{Roberto Turolla}
\affil{Dept. of Physics, University of Padova, \\ Via Marzolo 8, 35131 Padova,
Italy \\ e--mail: turolla@astaxp.pd.infn.it}
\author{Silvia Zane, Aldo Treves}
\affil{International School for Advanced Studies, \\ Via Beirut 2-4, 34014
Trieste, Italy \\ e--mail: zane@sissa.it, treves@astmiu.uni.mi.astro.it}
\and 
\author{Andrei Illarionov}
\affil{P.N. Lebedev Physical Institute, Russian Academy of Science, \\
Profsoyuznaya Ul. 84/32, Moscow 117810, Russia \\ e--mail: 
illarion@dpc.asc.rssi.ru}
\begin{abstract}
As first shown by \markcite{sh70}\Sh (1970), a neutron star accreting close to
the Eddington limit must acquire a positive charge in order for electrons and
protons to move at the same speed. The resulting electrostatic field may
contribute to accelerating positrons produced near the star surface in
conjunction with the radiative force. We reconsider the balance between energy
gains and losses, including inverse Compton (IC), bremsstrahlung and
non--radiative scatterings. It is found that, even accounting for IC losses
only, the maximum positron energy never exceeds $\approx 400$ keV. The
electrostatic field alone may produce energies $\approx 50$ keV at most.  We
also show that Coulomb collisions and annihilation with accreting electrons 
severely limit the number of positrons that escape to infinity. 
\end{abstract}


\section{Introduction}

It was suggested a long time ago (\markcite{sh70}\Sh 1970; 
\markcite{mic72}Michel 1972; \markcite{at73}Anile, \& 
Treves 1972; \markcite{mrt74}Maraschi, Reina, \& Treves 1974) that accretion of 
a hydrogen plasma
onto a neutron star must necessarily imply the presence of an electrostatic
field $\cal E$ under stationary conditions.
The field $\cal E$ arises because the radiation drag, due to the outflowing
flux, acts efficiently only on electrons while practically all the mass is 
in protons. The electrostatic field compensates the
different acceleration of the two species and allows electrons and protons to 
move at the same speed. Obviously $\cal E$ is stronger the
larger is the deviation from free--fall, a situation which is expected 
when the luminosity approaches the Eddington limit. 

The typical values of the charge $Q$ associated with the
electric field are not large enough to influence the space--time around 
the compact object (\markcite{mic72}Michel 1972), but its presence may 
nevertheless have interesting astrophysical consequences. 
\markcite{sh70}\Sh (1970) proposed that such a system could act as an 
electrostatic 
accelerator for positrons, for which the electrostatic and radiative forces
have the same direction. 
Balancing energy gains and losses, due mainly to inverse Compton with soft 
ambient photons, he estimated that positrons may be accelerated to 
energies of a few tens of MeV. 

Although the ``\Sh Accelerator'' is definitely ingenious, it received 
little attention in the past, possibly because no obvious positron source
seemed available near neutron stars. However, 
positrons may be generated by the radioactive decay of
the neutron star crust elements subject to the bombardment
of accreting particles (\markcite{rtt74}Reina, Tarenghi, \& Treves 1974;
\markcite{bsw92}Bildsten, Shapiro, \& Wasserman 1992). Pair production may be 
also expected in accretion models 
with shocks (see e.g. \markcite{ss75}Shapiro, \& Salpeter 1975) 
or in hydrostatic atmospheres kept at mildly relativistic temperatures by 
comptonization (\markcite{tzct94}Turolla, {\it et al.\/} 1994; 
\markcite{ztt96}Zane, Turolla, \& Treves 1996).

Here we reconsider \Sha's idea, examining the role of cooling 
processes different from inverse Compton 
(bremsstrahlung, non--radiative scattering) and 
the possibility that accelerated positrons survive 
annihilation with atmospheric 
electrons, escaping to infinity. 

\section{The Electric Field}

Let us consider spherical accretion onto a neutron star ($M=1.4 \MS$, $\rs = 
10^6$ cm)
and for simplicity suppose that the accreting material is ionized hydrogen.
Neglecting general relativistic corrections, the emerging luminosity is 
related to the accretion rate by
\begin{equation}
L = {{GM\mdot}\over\rs}\, .
\end{equation}

When  $L$  approaches the Eddington limit
$L_E = 4\pi GMm_pc/\sigma_T$,
the flow dynamics is strongly affected by the radiative
force (see e.g \markcite{mrt74}\markcite{mrt78}Maraschi, Reina, \& Treves 
1974, 1978; \markcite{mil90}Miller 1990; \markcite{pm91}Park \& Miller 
1991; \markcite{ztt93}Zampieri, Turolla, \& Treves 1993, hereafter ZTT).
Owing to the reduced effective gravity, the
flow deviates from free--fall and, for $L\sim L_E$, a settling behaviour
appears close to the star surface, where the velocity decreases with radius.
Even a small deviation from free--fall, which is always present for a 
non--vanishing luminosity, implies that electrons and protons are 
subject to different accelerations so that an electrostatic field $\cal E$
is built up around the star (\markcite{sh70}\Sh 1970; 
\markcite{zn71}Zeldovich, \& Novikov 1971; \markcite{mic72}Michel 1972). 
Although, in principle, the same argument applies to spherical accretion onto a
black hole, the efficiency is so low ($\lesssim 10^{-4}$ in absence of 
magnetic fields, see e.g. 
\markcite{ntz91}Nobili, Turolla, \& Zampieri 1991) that departures from 
free--fall are really negligible for any reasonable value of the accretion rate.
The expression for $\cal E$ can be
easily derived considering the forces acting on the two species ($e,\, p$) 
\begin{equation}
F_e = -{{L_{co}\sigma_T}\over{4\pi c r^2}} + {{GMm_e}\over{r^2}} +
{\cal E}e
\end{equation}
and
\begin{equation}
F_p = {{GMm_p}\over{r^2}} - {\cal E}e\, ,
\end{equation}
where $L_{co}$ denotes the luminosity in the frame comoving with the accretion
flow. Equating the two accelerations we get
\begin{equation}
{\cal E} \simeq {{L_{co}\sigma_T}\over{4\pi c e r^2}}\simeq 200 
\left({{10^6}\over r}\right)^2 {L_{co}\over L_E}\, {\rm V/cm}\, .
\end{equation}
At large radii, where $L_{co}\sim L$, the previous expression coincides with 
the result obtained by \markcite{mrt74}\markcite{mrt78}Maraschi, Reina, \& 
Treves (1974, 1978) analyzing the flow equations for the two species
\begin{equation}
{\cal E} = {{L\sigma_T}\over{4\pi c e r^2}}
\exp\left(-{L\over L_E}{\rs\over r}\right)\, .
\end{equation}
An estimate of the largest potential difference can be easily obtained 
integrating equation (4) assuming $L_{co}\sim L=L_E$ 
\begin{equation}
(\Delta V)_{max}\sim {{GM m_p}\over{e\rs}}\simeq 200 \ {\rm MV}
\end{equation}
which shows that, in principle, extremely relativistic positrons may be 
expected to reach infinity. 

\section{Compton Scattering}

Here we consider a scenario in which the number density of 
$e^+$ is not zero at some radius $r_{in}$. Positrons are 
immediately accelerated by the electric field and gain energy at a rate
\begin{equation}
\dot E_{\cal E} = ve{\cal E} \, , 
\end{equation}
where $v$ is the particle velocity.
At the same time, positrons are accelerated by the radiative force and 
suffer radiative losses through inverse Compton (IC) scatterings with the
radiation bath; $e^+-p$ and $e^+-e^-$ 
bremsstrahlung; non--radiative scatterings with the flow electrons.
In this section we concentrate on IC losses, which is the only 
cooling process taken into account by \Sha, and we 
show that, even in this simple case, ultra--relativistic energies are 
not expected to be reached. In the next section we will verify that 
bremsstrahlung losses are negligible, but that elastic scatterings may be 
very efficient in transfering energy to ambient electrons and will ultimately 
stop the positron flow. 

The energy lost by a positron for IC scattering with a radiation 
field of energy density $U$ and radiation pressure $K$ is 
\begin{equation}
\dot E_{IC} = - c\sigma_T \gamma^2 \beta^2 (U + K)\, , 
\end{equation}
while the gain due to the radiative force is
\begin{equation}
\dot E_{rad} = c\sigma_T \gamma^2 \beta (1 + \beta^2 ) F\, , 
\end{equation}
where $F$ is the total flux divided by $c$ 
(\markcite{gr65}Gurevich, \& Rumyantsev 1965, all the radiation moments are 
here evaluated in the lab--frame). 
For an isotropic radiation field, $F = 0$, $K = U/3$ 
and the expression for the net power reduces to 
$\dot E_{IC} = - 4 
c\sigma_T U \gamma^2 \beta^2/3  $ (see e.g. \markcite{rl79}Rybicki, \& 
Lightman 1979). 
Accordingly, the positron energy changes 
with radius as
\begin{equation}
{{dE}\over{dr}} = e{\cal E} +
\sigma_T \gamma^2 ( 1 + \beta^2 ) F
- \sigma_T (U+ K) \gamma^2\beta\, .
\end{equation}

Introducing the dimensionless quantities $x=r/r_g$, 
$u = 4\pi r_g^2c U/L_E$, 
$k = 4\pi r_g^2c K/L_E$, 
and $l = L/L_E$, where $r_g$ is the Schwarzschild radius, equation (10)
can be written as
\begin{eqnarray}
{{d\gamma}\over{dx}} = {{m_p}\over{2m_e}}{ 1\over{x^2}}\left[l_{co}+
(2\gamma^2-1)l -\right.\nonumber \\
\left. x^2(u+ k)\gamma\sqrt{\gamma^2-1}\right]\, .
\end{eqnarray}

Whenever $x\ll x_c = m_p/2m_e$ the solution of equation (11) relaxes 
immediately to the reduced solution 
\begin{equation}
\beta = {{x^2(u + k)}\over{2\left(l_{co}-l\right)}}\left[\sqrt{1+ 
{{4\left(l_{co}^2-l^2\right)}\over{
x^4(u+k)^2}}}-1\right]
\end{equation}
irrespective of the initial particle energy.
If no other processes
prevent positrons from escaping (see the discussion in section 4), the 
terminal Lorentz factor can be obtained integrating numerically 
equation (11) once the radiation field in the accreting medium is known. 
In section 5 numerical results based on the accretion model described in
ZTT will be presented. 

Equation (12) also allows a close comparison between our result and \Sha's
one. The contribution of electrostatic acceleration can be easily
obtained dropping $\dot E_{rad}$ in equation (10) which is, formally,
equivalent to put $l=0$ in equation (12). An upper limit for the terminal 
Lorentz factor in this case is given by
$\gamma_\infty^2 = \left(1+\sqrt{2}\right)/2\simeq 1.2$, 
which corresponds to a kinetic energy $\sim 50$ keV. 
This is much less than the value of tens of MeV proposed by \Sha.
The point in \Sha's original analysis that should be reconsidered 
is the calculation of the IC
losses, which were estimated in a very crude way by multiplying the energy
exchange in each collision times an ``optical depth'' parameter, $\tau_+$. 
Even assuming that the total number of collisions suffered by the positron in 
traveling a radial distance $r$, $N$,
is indeed related to \Sha's equation (2) (and this is questionable), $N$ should
be $\approx\tau_+^2$, since $\tau_+\gg 1$. This implies that \Sha's 
Compton cooling is underestimated by a factor $\approx\tau_+\sim 10^5$. Moreover
the final particle energy is evaluated assuming that no losses are present 
for $\tau_+<1$ which results in a even larger underestimate of IC cooling. 
 
\section{Other Interactions and Relevant Timescales}

In the previous section we have derived the energy equation for positrons,
including only IC cooling. 
In section 5 we show that, even under this favourable assumption, 
the positrons terminal Lorentz factor is $\lesssim 2$ (see table
1). In order to establish if indeed these 
mildly relativistic particles may reach infinity, we need a more thorough 
understanding of the physical conditions in the flow close to the star 
surface. In our scenario four different kinds of particles are 
present: ``primary'' electrons and protons of the accreting 
material, with the same number 
density $n$ and temperature $T$, together with positrons and electrons,  
with number density $n_+$ and temperature $T_+$, that are created in the 
atmosphere by photon--photon or particle collisions. We will limit our 
discussion to the case in which $z = n_+ / n \ll 1$, so the number 
density of electrons is $n_- = n + n_+ \sim n$. 
The first issue that should be addressed concerns the probability 
that positrons annihilate with ambient electrons. This question arises quite 
naturally since, at variance with 
\Sha's scenario in which highly relativistic energies are expected, 
now the particle energy is not high enough to ensure that the 
cross section for pair annihilation becomes negligible.        

In the non--relativistic limit, the annihilation rate for positrons 
and electrons with arbitrary 
distribution functions is (\markcite{sv82}Svensson 1982)
\begin{equation}
\dot n_+  = n_+ n_- c \pi r_e^2 \, . 
\end{equation}
We introduce a characteristic time $t_{A}$ for annihilation defined as  
\begin{equation}
t_{A} = {n_+\over \dot n_+}
= {1 \over n_- c \pi r_e^2}  
\sim 1.34 \times 10^{-3} \left ( { 10^{17} \over n} \right ) \  
{ \rm s} \, . 
\end{equation}
Since $t_{A}$ depends on $n$, it gives 
the probability of annihilation on a scale of order the density lengthscale 
in the accretion flow, $h\sim r$.
On the other hand, a positron of energy $\gamma$ covers a distance  
$h$ in a time $t_d$ 
\begin{equation}
t_d = \int_r^{r+ h} 
{ \gamma \over c \sqrt { \gamma^2 - 1 }}  d r
\simeq \left . { \gamma  \over c \sqrt { \gamma^2 - 1 }} \right |_{r+h/2} h
\, , 
\end{equation}
and the particle survives annihilation if $t_{d} < t_{A}$.  
A numerical comparison between these two timescales will be given in 
the next section for different values of both the emergent luminosity and 
$r_{in}$. Here we only note that, in the more favourable situations in 
which particles are immediately accelerated up to $\gamma\sim 1.7$, 
this condition is met if 
\begin{equation}
n < { 1 \over \pi r_e^2} { \sqrt { \gamma^2 - 1} \over \gamma r } 
\simeq  3.2 \times 10^{18} \left ( { 10^6\over r }\right ) \ {\rm cm^{-3}} 
\, . 
\end{equation}
This limit is fairly easy to satisfy for near Eddington accretion since 
$n = \mdot/(4\pi r^2m_pv)$$\sim L_E/(4\pi r^2m_pc^3)$$\approx 10^{17}
(10^6/r)^2 \ {\rm cm}^{-3}$ close to the star surface.
 
Other cooling mechanisms that could influence the maximum 
positron energy are radiative and non--radiative scattering with ions and 
electrons in the accretion flow. 
The positron--ion bremsstrahlung energy 
loss is given by 
\begin{equation}
\dot E_{B} \simeq - 10^{-26} Z^2 n c E \ {\rm erg\, s}^{-1}\, , 
\end{equation}
where $Z$ is the charge of the ions. 
This process acts on a characteristic timescale 
\begin{equation}
t_{B} = {E \over \dot E_{B}}
\simeq 3.3 \times 10^{-2} Z^{-2} \left ({ 10^{17} \over n } \right 
) \ { \rm s} \, ,
\end{equation}
which, compared with $t_{A}$, yields
\begin{equation}
{t_{A} \over t_{B}} \simeq 4 \times 10^{-2} Z^2\, , 
\end{equation}
so, for pure hydrogen, 
bremsstrahlung losses are negligible 
for particles surviving annihilation and free--free could 
represent a competitive cooling process only for large $Z$.

On the other hand, the energy exchange due to collisions is indeed important
and can modify substantially the picture outlined up to now.
Let us introduce the energy loss rate $\nu_E$ for
a test particle in a plasma of protons and electrons as
\begin{equation}
\dot E_{el} =  - \nu_E E \, , 
\end{equation}
and focus on the non--relativistic regime. 
The more efficient process is radiationless, elastic 
scattering between positrons and electrons; positron--proton
equipartition needs
a larger timescale by a factor $m_p/m_e$. The energy loss rate can be 
obtained from the expression of diffusion coefficients, which, in the 
limit $v$ much larger than the proton velocity, yields 
(see e.g. \markcite{rs83}Rosenbluth, \& Sagdeev 1983)
\begin{eqnarray}
\nu_E \sim { 8 \pi  e^4 n \ln \Lambda  \over m_e^2 v^3 } 
\left [ {m_e \over m_p } + {\rm erf} \left ( {v \over v_e  } \right )\right. - 
\nonumber \\ 
\left.{4 \over \sqrt{\pi}  }{v \over v_e  }e^{-\left ( v / v_e   \right )^2} 
\right ] \, , 
\end{eqnarray}
where $\ln \Lambda $ is the Coulomb logarithm and $v_e$ the electron velocity.  
The rate $\nu_E$ is not necessarily positive since a fast particle tends 
to lose energy to the plasma and a slow one to gain energy.
Initial  differences in the mean velocities of positrons and electrons 
are eliminated by collisions in a time 
$t_{coll} \sim \left | \nu_E \right |^{-1}$.
The importance of collisions can be quantified 
comparing $t_{coll}$ with the dynamical timescale given by expression 
(15).

\section{Numerical Estimates}

In order to investigate the effects of electrostatic plus radiative
acceleration on  particles injected in different atmospheric layers, 
we have 
solved equation (11) varying both the accretion  luminosity and the 
injection radius $r_{in}$. The run of the flow variables has been taken 
from the one--fluid models computed by ZTT for near--Eddington accretion of a
``cold'', pure scattering medium onto neutron stars.
The solutions are plotted, together with 
the reduced solution (equation 12), in figures 1a, b, c for $L/L_E =$0.3, 
0.6 and 0.9, respectively. For any given luminosity, we have considered 
three different values of $r_{in}$, which correspond to a scattering 
optical depth $\tau \sim 0.1, 1, 5$ (see table 1), with the exception 
of the case $L/L_E = 0.3$, for which the maximum optical depth is 
$\tau \simeq 1.5$ at the star surface. This simple choice allows us to 
mimic the acceleration of positrons initially created in 
different atmospheric layers, and, increasing the luminosity, to visualize 
the process in the presence of more and more expanded atmospheres. In order 
to account for the anisotropy of the radiation field, we introduced the 
closure $f(\tau) = K/U$, adopting for $f$ the same 
expression used by ZTT, $f(\tau) = 2/\left [ 3 \left ( 1 + \tau^2 \right 
) \right ] + 1 /3$. The final energies in all the cases we considered 
are listed in table 1. Figures 2a, b, c show the comparison between 
$t_d$ and $t_A$ for the same choice of the parameters.
As can be seen from the figures, the reduced solution is always attained 
at small radii, but at the same time annihilation acts efficiently in 
the inner dense atmospheric layers preventing positrons created in optically 
thick regions from escaping. The chance to surviving annihilation increases for 
particles created in photospheric regions and above, 
although positrons starting at $\tau \lesssim 1$ reach infinity with a lower
Lorentz factor. 
The larger terminal velocities are expected for low values of $L$
when the photosphere is closer to the star surface.    

Figures 3a, b, c show $t_{d}$ together with $t_{coll}$ for 
two different values of the gas temperature, since the efficiency of 
Coulomb collisions strongly depends on the difference between the test 
particle and the electron velocities. We note that only for $kT\gtrsim 70$ 
keV is $\nu_e$ negative in an extended region near $r_{in}$, so the
possibility that positrons gain additional energy owing to collisions
appears remote. 
On the other hand, under the more realistic assumption that $kT \lesssim 10$ 
keV, as expected in typical X--ray sources, we find that mildly 
relativistic positrons are ``de facto'' prevented from reaching infinity,
all differences between the mean energies of the two species being
washed out by the collective effect of soft collisions.

\section{Discussion}

In this paper we have reconsidered acceleration of positrons near a
neutron star accreting close to the Eddington limit. The 
radiative force and the electrostatic 
thrust due to the non--vanishing electric field around the neutron star 
are opposed by 
inverse Compton losses and elastic scattering with electrons of the 
accreting material. Ultimately, these are the main agents
responsible for both the acceleration and deceleration of positrons.

The maximum energy which can be achieved by positrons 
owing to the electrostatic acceleration alone is limited to $\sim 50$ keV
which differs substantially from \Sha's estimate of few 
tens of MeV.
The main contribution to the particle's acceleration is therefore provided 
by the radiative force, which acts equally on electrons and positrons.
We have also shown that the largest 
terminal Lorentz factors are still $\lesssim 2$, even if this additional 
gain is accounted for. A first consequence is 
that the maximum energies attained by positrons are in a range  
where pair annihilation is still important, at least 
for particles created at $\tau > 1$. Moreover, elastic scatterings with 
cold electrons in the accretion flow oppose the particle acceleration 
and provide an additional source of trapping. This suggests that 
only a fraction of positrons will actually leave the star, while the rest 
should annihilate and eventually contribute to a narrow 511 keV line.

Furthermore, no multiplication through electromagnetic 
showers either on the accreting material or on the photons can be 
expected, so the positron production depends
mainly on the injection process. Moreover, we note that the same  
conclusions discussed above can be applied to particles injected below 
$\sim 400R_*$ (i.e. below $x_c$), quite independently of the 
initial conditions. If the source is close to the Eddington 
limit, even if positrons are injected with energies $\sim$ MeV, like in 
radioactive decay, because of the efficiency of IC 
losses, their final energy will be limited to $\sim 400$ keV and 
the possibility of outstream or annihilation will not change.

\clearpage

\begin{deluxetable}{ccccc}
\tablecolumns{5}
\tablenum{1}
\tablecaption{The radius, the proton number density and terminal kinetic 
energy for different values of the accretion  luminosity and of the 
scattering optical depth at the injection point\label{tab1}}
\tablehead{
\colhead{$L/L_E$} &
\colhead{$\tau$}&
\colhead{$r_{in}$} &
\colhead{$n(r_{in})$} &
\colhead{$E_\infty$} \\
\colhead{} &
\colhead{} &
\colhead{$10^6$ cm} &
\colhead{$ 10^{17}$ cm$^{-3}$} &
\colhead{keV}}
\startdata
 0.3 & 1.5 &   1 & 19    & 358 \nl
 0.3 & 1.  &   2 &  5    & 358 \nl
 0.3 & 0.1 & 141 &  0.006& 296 \nl
 0.6 & 5.  & 1.5 & 61    & 293 \nl
 0.6 & 1.  &  13 &  0.5  & 293 \nl
 0.6 & 0.1 & 855 &  0.002& 151 \nl
 0.9 & 5.  &  11 &  5    & 198 \nl
 0.9 & 1.  & 123 &  0.06 & 198 \nl
 0.9 & 0.1 &5000 & 0.0002&  50 \nl
\enddata
\end{deluxetable}

\clearpage

\figcaption[elac1.eps]{a) The positron energy vs. radius
for $L/L_E = 0.3$ and scattering depth at injection 1.5, 1, and  0.1.
The reduced solution is also shown (dashed line);
b) same as in figure 1a for $L/L_E = 0.6$ and 
$\tau = 5$, 1 and 0.1; c) same as in figure 1b for $L/L_E = 0.9$.\label{fig1}}

\figcaption[elac2.eps]{The annihilation time $t_A$ 
(dashed line) and the dynamical time $t_d$ (solid lines) 
for the same values of the luminosity and injection depth as in figure 1.
\label{fig2}}

\figcaption[elac3.eps]{The collisional time $t_{coll}$ for $kT = 10, 70$ keV 
(dashed
lines) and the dynamical time $t_d$ (solid line) for $L/L_E = 0.3$ and 
$\tau = 1$; b) same as in figure 3a for $L/L_E = 0.6$; c) same as in figure 
3a for $L/L_E = 0.9$.\label{fig3}} 

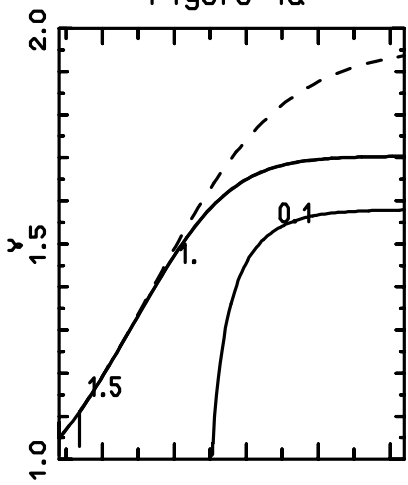
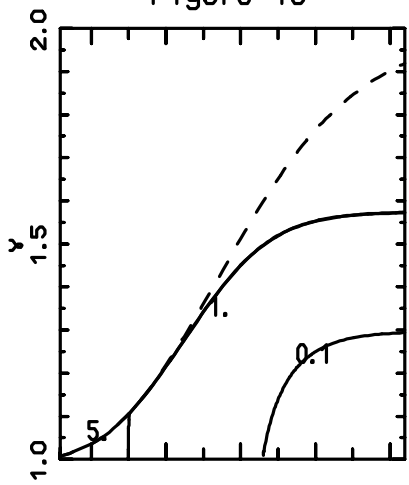
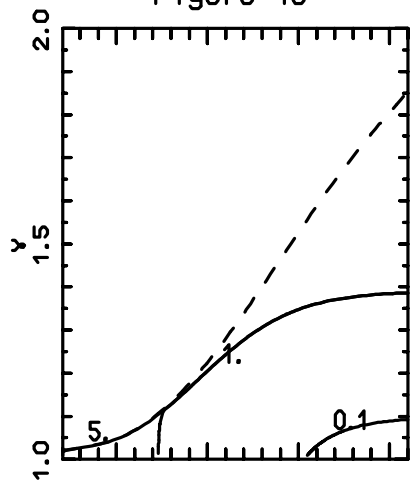
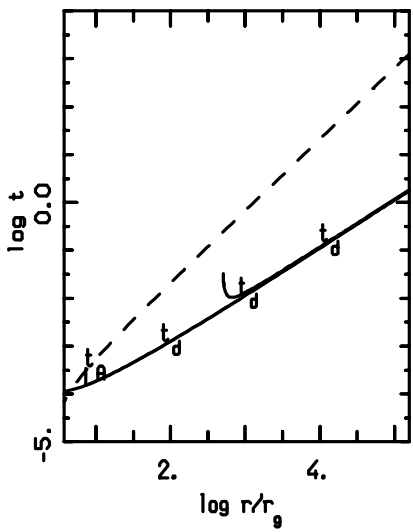
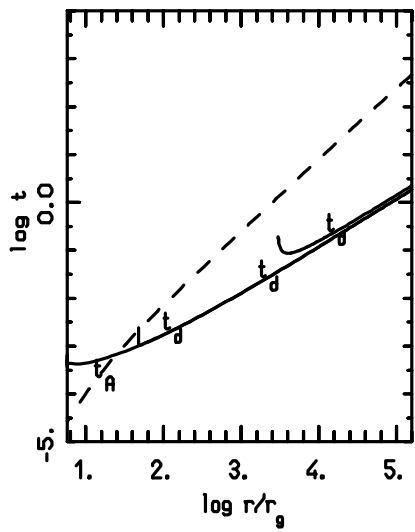
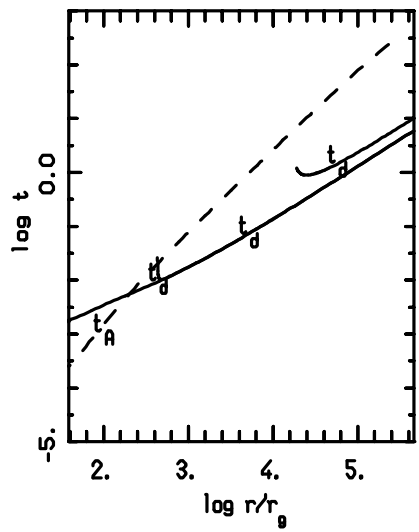
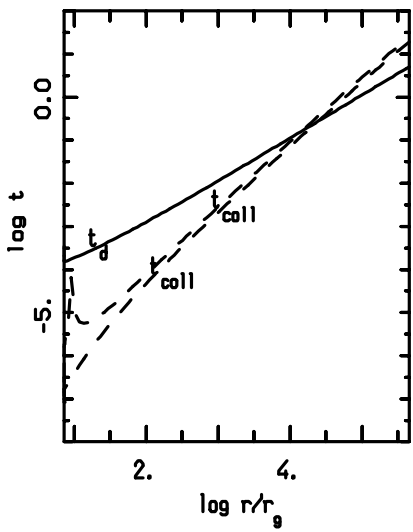
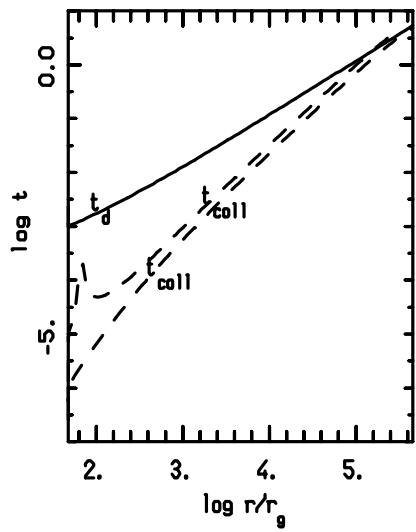
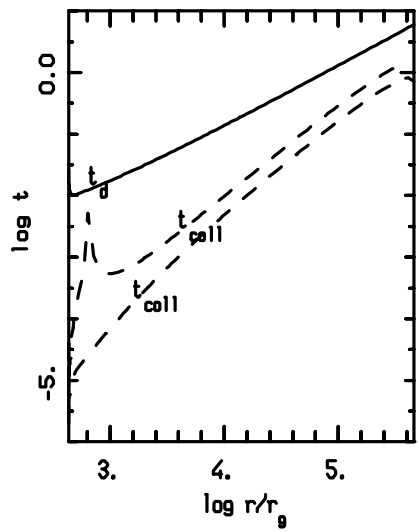
\end{document}